\documentclass[11pt,preprintnumbers,superscriptaddress,amsmath,amssymb,aps,nofootinbib]{revtex4}
\usepackage{graphicx}% Include figure files
\usepackage{dcolumn}% Align table columns on decimal point
\usepackage{bm}% bold math
\usepackage{amssymb}
\usepackage{amsmath}
\usepackage{epsfig}
\usepackage{color}
\usepackage{slashed}
\usepackage{hhline}
\usepackage{ulem}
\usepackage{bbm}
\usepackage{tikz-feynman}
\tikzfeynmanset{compat=1.1.0}
\def\be{\begin{equation}}
\def\ee{\end{equation}}
\newcommand{\bea}{\begin{eqnarray}}
\newcommand{\eea}{\end{eqnarray}}
\newcommand{\nn}{\nonumber}

\newcommand{\ecm}{e \, {\rm cm}}

\begin{document}

\title{Radiative lepton model in a noninvertible fusion rule}

\author{Takaaki Nomura}
\email{nomura@scu.edu.cn}
\affiliation{College of Physics, Sichuan University, Chengdu 610065, China}

\author{Hiroshi Okada}
\email{hiroshi3okada@htu.edu.cn}
\affiliation{Department of Physics, Henan Normal University, Xinxiang 453007, China}

\author{Yoshihiro Shigekami}
\email{shigekami@htu.edu.cn}
\affiliation{Department of Physics, Henan Normal University, Xinxiang 453007, China}

\date{\today}

\begin{abstract}
We propose a radiatively induced lepton mass model introducing a $Z_2$ gauging $Z_5$ fusion rule. 
In our framework, the charged-lepton mass matrix is generated at one-loop level via dynamical breaking of the fusion rule. 
On the other hand, the neutrino mass matrix is induced at the one-loop level without breaking the fusion rule. 
As a direct consequence of the loop induced charged lepton masses, we can also consider lepton flavor violations, electron and muon $g-2$, and charged-lepton electric dipole moments that come into our valid phenomenological discussion. 
Then, we perform numerical analysis and show some interesting tendency on the Dirac $CP$ phase, two Majorana phases, charged-lepton electric dipole moments and the neutrinoless double beta decay, all of which depends on their arguments where we fix the absolute values of our free parameters in order to satisfy experimental data of the lepton masses and mixing angles. 
\end{abstract}

\maketitle

\newpage

\section{Introduction}

The radiative seesaw model is one of the attractive scenarios to understand a mechanism of minuscule masses such as active neutrinos and electrons at low energy scale beyond the standard model (BSM). 
In addition, we can potentially discuss plenty of phenomenology such as lepton flavor violations (LFVs), electron/muon anomalous magnetic dipole moments (electron/muon $g-2$), charged-lepton electric dipole moments (EDMs), and dark matter (DM), all of which can be connected to such tiny mass particles inside the loop. 
To obtain relevant terms in order to induce loop diagrams as leading contributions,
noninvertible fusion rules are recently applied to these kinds of models. 
These selection rules possess some intriguing natures. 
The first remarkable nature is that the rules can forbid undesired terms at tree level but violate at loop level in a dynamical manner. 
The second one is that these rules provide an exact symmetry such as $Z_2$ that is never broken at any orders and plays a role in stabilizing the DM candidate. 
The third one is that they give some mass textures to reduce mass parameters in quark and lepton sectors, and possibly lead to predictions.\footnote{See Refs.~\cite{Jiang:2025psz,Kobayashi:2025rpx,Jangid:2025krp,Okada:2025kfm,Choi:2022jqy,Cordova:2022fhg,Cordova:2022ieu,Cordova:2024ypu,Kobayashi:2024cvp,Kobayashi:2024yqq,Kobayashi:2025znw,Suzuki:2025oov,Liang:2025dkm,Kobayashi:2025ldi,Kobayashi:2025lar,Kobayashi:2025cwx,Nomura:2025sod,Dong:2025jra,Nomura:2025yoa,Chen:2025awz,Kobayashi:2025thd,Suzuki:2025bxg}, which proposed several ways to apply these rules for phenomenology.} 
Note here that the second and third nature can be replaced by the other symmetries such as discrete flavor symmetries~\cite{Altarelli:2010gt,Ishimori:2010au,Kobayashi:2022moq,Ishimori:2012zz}. 
However, the first one is totally unique for the case of the noninvertible fusion rule. 
These natures pave the way for constructing new models and it is worth exploring applications of the rules. 

In this work, we apply $Z_2$ gauging $Z_5$ fusion rule ${\bf Z_5^{NI}}$, which consists of three elements $\{ \mathbbm{I}, a, b\}$, where all are supposed to be real and commutable each other; backgrounds and rules for $Z_2$ gauging $Z_N$ symmetry can be referred to Refs.~\cite{Kobayashi:2024yqq,Kobayashi:2024cvp}. 
For these elements, we have
\begin{align}
a \otimes a = \mathbbm{I} \oplus b, \quad a \otimes b = a \oplus b, \quad b \otimes b = \mathbbm{I} \oplus a, \quad \mathbbm{I} \otimes (\mathbbm{I}, a, b) = (\mathbbm{I}, a, b).
\label{eq:Z5NIrules}
\end{align}
Here, it is worthwhile to comment on the mathematical nature of our noninvertible symmetry. 
Since all the generators $\mathbbm{I},\ a,\ b$, mutually commute, one might naively expect that ${\bf Z_5^{NI}}$ shares the characteristics of Abelian group symmetries such as $Z_N$ or $U(1)$. 
However, the multiplication rules given in Eq.~\eqref{eq:Z5NIrules} exhibit a structure reminiscent of non-Abelian groups. 
Nevertheless, the algebraic structure of ${\bf Z_5^{NI}}$ differs fundamentally from that of non-Abelian groups. 
In an ordinary Abelian group, the product of two arbitrary generators always results in a single generator, i.e., $a \otimes b = c$. 
Therefore, the fusion rule in Eq.~(\ref{eq:Z5NIrules}) cannot be realized within the framework of conventional Abelian groups. 
Consequently, ${\bf Z_5^{NI}}$ provides a genuinely new structure that is not reducible to any standard group symmetry.\footnote{These features can be generalized to other noninvertible symmetries.} 
Owing to these distinctive features, it opens up the possibility of constructing novel scenarios that cannot be achieved using conventional group-theoretic symmetries. In the main text, we explicitly demonstrate these differences in detail. 
For a radiative lepton mass model, we introduce left- and right-handed Majorana fermions, isospin doublet inert boson $\eta$, singly charged-boson $S^-$.\footnote{We can also realize the same interactions by applying Fibonacci fusion rules that is the minimum fusion ones. 
However, this scenario induces terms that are forbidden at tree level, by radiative corrections. 
Some of these radiative corrections include infinity that never be able to be canceled out. 
We thus do not adopt the minimal Fibonacci fusion rules to keep predictability of the model. 
We briefly show some more detail in the model part.} 
We make use of the first and second nature in the ${\bf Z_5^{NI}}$ fusion rules. 
Therefore, the charged-lepton mass matrix can be induced at one loop via dynamical breaking of ${\bf Z_5^{NI}}$. 
On the other hand, the neutrino mass matrix is generated at the one-loop level where ${\bf Z_5^{NI}}$ provides an accidental $Z_2$ symmetry that never can be broken at any order. 
This is the same as the case of Ma model~\cite{Tao:1996vb,Ma:2006km}. 
In addition, through the charged-lepton sector, we can also discuss LFVs, electron/muon $g-2$ and charged-lepton EDMs.\footnote{Even though we can discuss the nature of DM candidate, we focus on the other phenomenology such as EDMs in this paper.} 

This paper is organized as follows. 
In Sec.~\ref{sec:II}, we review our lepton seesaw model and explain how to induce the small charged-lepton masses as well as the neutrino masses. 
We also discuss several phenomenology related to the charged leptons. 
In Sec.~\ref{sec:III}, we demonstrate several numerical $\chi^2$ results satisfying neutrino observables, LFVs, electron/muon $g-2$, and charged-lepton EDMs in cases of normal and inverted hierarchies. 
In Sec.~\ref{sec:IV}, we summarize and conclude.

\section{Model setup}
\label{sec:II}

Here, we review our model setup in which we introduce $Z_2$ gauging $Z_5$ fusion rule ${\bf Z_5^{NI}}$. 
First, the lepton masses at tree level are forbidden by assigning ${\bf Z_5^{NI}}$ charge to lepton particles. 
Then, in order to generate the lepton mass matrix at the one-loop level, we introduce three families of left/right neutral Majorana fermions $N_{L/R}$, an isospin doublet inert boson $\eta \equiv [\eta^+,(\eta_R+i\eta_I)/\sqrt2]^T$ and an isospin singlet singly-charged-boson $S^-$ in addition to the SM particles. 
Moreover, $\eta$ and $S^-$ do not acquire nonzero vacuum expectation values (VEVs). 
Note here that the charged-lepton mass matrix is generated by dynamically breaking of this symmetry,~\footnote{It means noninvertible symmetry is broken by radiative corrections where charged lepton mass is forbidden at tree level by the symmetry. 
Note that effective operator inducing neutrino mass, $LLHH$, is allowed by the symmetry. 
These features can be explicitly observed from diagrams for the radiative lepton masses in Figs.~\ref{fig:ClepMass} and \ref{fig:NuMass}.} and LFVs, $g-2$, and EDMs have the same origin as the charged-lepton mass matrix~\cite{Khaw:2022qxh, Okada:2013iba, Okada:2014nsa}. 
On the other hand the neutrino mass matrix is obtained without violating the fusion rule, as we will show later. 
The particle contents and their charge assignments are summarized in Table~\ref{tab:1}. 
\begin{table}[t!]
\begin{tabular}{|c||c|c|c|c||c|c|c|}\hline\hline 
 & ~$L_L$~ & ~$\ell_R$~ & ~${N_L}$~ & ~${N_R}$~ & ~$H$~ & ~{$\eta$}~ & ~{$S^-$}~ \\\hline\hline
$SU(2)_L$ & $\bm{2}$ & $\bm{1}$ & $\bm{1}$ & $\bm{1}$ & $\bm{2}$ & $\bm{2}$ & $\bm{1}$ \\\hline
$U(1)_Y$ & $-\frac12$ & $-1$ & $0$ & $0$ & $\frac12$ & $\frac12$ & $-1$ \\\hline
${\bf Z_5^{NI}}$ & ${b}$ & $a$ & $a$ & $a$ & $\mathbb{I}$ & $ b$ & $b$ \\\hline
\end{tabular}
\caption{Charge assignments of relevant fermions and bosons
under $SU(2)_L\otimes U(1)_Y \otimes {\bf Z_5^{NI}}$. 
All fermions have to have three generations, and all new particles are singlet under $SU(3)_c$. }
\label{tab:1}
\end{table}

Under the symmetry, the renormalizable Lagrangian is given by
\begin{align}
- {\cal L}_{\ell} &= y^{\ell}_{ii} \overline{L_{L_i}} \eta \ell_{R_i} + f^R_{ia} \overline{L_{L_i}} {\tilde \eta} N_{R_a} + f^L_{ia} \overline{L_{L_i}} {\tilde \eta} N_{L_a}^C + g^R_{ai} \overline{N_{R_a}^C} \ell_{R_i} S^+ + g^L_{ai} \overline{N_{L_a}} \ell_{R_i} S^+ \nn \\
&\hspace{1.2em} + (M_D)_{aa} \overline{N_{L_a}} N_{R_a} + (M_R)_{ab} \overline{N_{R_a}^C} N_{R_b} + (M_L)_{ab} \overline{N_{L_a}} N_{L_b}^C+{\rm H.c.},
\label{eq:lpy}
\end{align}
where $\tilde\eta\equiv i\tau_2 \eta^*$ being $\tau_2$ the second Pauli matrix. 
$y^{\ell}$ and $M_D$ can be diagonal matrices without loss of generality. 
The Higgs potential is given by
\begin{align}
{\cal V} &= - \mu_H^2 |H|^2 + \mu_{\eta}^2 |\eta|^2 + \mu_S^2 |S^-|^2 + \mu_0 \left( H^T (i \tau_2) \eta S^- + {\rm H.c.} \right) \nn \\
&\hspace{1.2em} + \lambda_H |H|^4 + \lambda_{\eta} |\eta|^4 + \lambda_S |S^-|^4 + \lambda_{H \eta} |H|^2 |\eta|^2 + \lambda'_{H \eta} |H^{\dag} \eta|^2 + \lambda''_{H \eta} \left( (H^{\dag} \eta)^2 + {\rm H.c.} \right) \nn \\
&\hspace{1.2em} + \lambda_{H S} |H|^2 |S^-|^2 + \lambda_{\eta S} |\eta|^2 |S^-|^2 .
\label{eq:vp}
\end{align}
Especially, the $\lambda''_{H \eta}$ term contributes to generating the neutrino mass matrix at one-loop level, and the $\mu_0$ term contributes to the radiative charged-lepton mass matrix. 

Before moving to see the radiative mass generation for each fermion sector, we comment on the physical phases of the model. 
In the Higgs potential in Eq.~\eqref{eq:vp}, there are two possible complex parameters, $\mu_0$ and $\lambda''_{H \eta}$. 
These phases can be absorbed into the two phases of $H$, $\eta$, and $S^{\pm}$. 
Since other parameters in the Higgs potential are real by definition, there is no physical $CP$ source from the potential. 
In the Lagrangian in Eq.~\eqref{eq:lpy}, on the other hand, we have lots of complex parameters. 
It is clear that all these phases cannot be removed by considering phase redefinition for remaining fields. 
As a result, we truly have physical $CP$ sources which result in nonzero (and nontrivial) predictions for charged-lepton EDMs. 
For example, following phases are invariant under any rephasing for fields:
\begin{align}
\theta_{\rm phys} = \arg \left[ \frac{f^R_{ia} g^L_{ia}}{f^L_{ia} g^R_{ia}} \right], ~ \arg \left[ \frac{(M_D)_{aa}^2}{(M_R)_{aa} (M_L)_{aa}} \right], ~ \cdots,
\label{eq:CPexample}
\end{align}
with no summation for $i, a$. 

Furthermore, we emphasize the role of noninvertible symmetry. 
If we use ordinary global symmetry, like $Z_M$ or global $U(1)$ symmetry, it is not possible to realize the scenario as explained below. 
Let us consider $Z_4$ symmetry is used. 
First, we need to assign $Z_4$ charge ``$-1$" for the inert scalar doublet $\eta$ in any setup, in order to have the $(H^{\dagger} \eta)^2$ term,\footnote{For this term, it is clear that we cannot use any discrete $Z_M$ symmetry with odd number of $M$.} which is necessary for the neutrino mass generation (SM Higgs should not have charge to generate quark mass). 
Then $S^-$ should also have $Z_4$ charge $-1$ for the $H \eta S^-$ term. 
In this case, we can write terms $\{ \overline{L_L} \eta \ell_R, \overline{L_L} \tilde{\eta} N_R, \overline{N^C_R} \ell_R S^+ \}$ by assigning charge ``$i$" for $\{\overline{L_L}, N_{L(R)}, \ell_R\}$. 
However, we cannot write Majorana mass terms of $N_{L(R)}$, and our one-loop diagrams generating lepton masses cannot be written. 
As an another option, we can assign $-1$ charge for $\overline{L_L}$ as well as $\eta$ and $S^{\pm}$, while $\ell_R$ and $N_{L(R)}$ do not have charge. 
In this case, we cannot write the $\overline{N^C_R} \ell_R S^+$ term, and charged lepton masses are not generated. 
These difficulties cannot be improved even if we use a simple $Z_2$, larger $Z_M$, or global $U(1)$ symmetry, and therefore, such a usual symmetry cannot be used for our scenario. 
By using the noninvertible fusion rule, on the other hand, we can realize it in use of the rules in Eq.~\eqref{eq:Z5NIrules}. 
Therefore our model is a good example to realize the scenario that cannot be obtained from the usual symmetry. 
Indeed, we have many free parameters, since we need several Yukawa couplings to induce charged lepton and neutrino masses at loop level. 
However, our main point in this paper is to construct a model in which all lepton masses are generated by radiative corrections rather than getting specific predictions. 

Notice that if we use the Fibonacci rules, $1 \otimes \tau = \tau$ and $\tau \otimes \tau = 1 \oplus \tau$, assigning $\tau$ to $\eta$ and $S^-$, we can write the term $(H^{\dagger} \eta) (\eta^{\dagger} \eta)$. 
This term can induce terms such as $(H^{\dagger} \eta) (H^{\dagger} H)$ and $\eta^{\dagger} H$ through radiative correction. 
These loop corrections generate divergences that cannot be absorbed by renormalization, since the corresponding tree-level counterterms are absent. Consequently, the divergences cannot be eliminated unless a finite cutoff scale is introduced. 
In addition, $\eta^{\dagger} H$ term spoils inert nature of $\eta$ and the mechanism of the model does not work. 
Therefore we need ${\bf Z_5^{NI}}$ as a relevant minimal selection rule where $(H^{\dagger} \eta)(\eta^{\dagger} \eta)$ is forbidden as $b \otimes b \otimes b$ does not contain $\mathbbm{I}$.

\subsection{Neutral fermion mass matrix}

The mass matrix of the neutral fermion is induced via the second line of Lagrangian in Eq.~\eqref{eq:lpy}, and its form is given by
\begin{align}
\left( \begin{array}{cc}
\overline{N^C_R} & \overline{N_L} 
\end{array} \right) \left( \begin{array}{cc}
M_R & M_D \\
M_D &M_L
\end{array} \right) \left( \begin{array}{c}
N_R \\
N^C_L
\end{array} \right) \equiv \left( \begin{array}{cc}
\overline{N^C_R} & \overline{N_L}
\end{array} \right) M_N \left( \begin{array}{c}
N_R \\
N^C_L
\end{array} \right),
\label{eq:cgd-mtrx}
\end{align}
where we omit generation indices, and each block is understood as a $3 \times 3$ matrix. 
$M_N$ is then understood as a $6 \times 6$ symmetric matrix. 
Note that due to our assumption about absence of VEVs for $\{ \eta, S^- \}$, neutrinos do not mix with $N_{L, R}$. 
In order to diagonalize the above mass matrix, we suppose
\begin{align}
M_R, M_L \ll M_D,
\end{align}
which is an appropriate hypothesis when we consider the `t Hooft sense. 
Then, we first transform $M_N$ as follows:
\begin{align}
M'_N \equiv V^{(1) T}_N M_N V^{(1)}_N& \equiv \left( \begin{array}{cc}
\frac{1}{\sqrt2} & \frac{1}{\sqrt2} \\
- \frac{1}{\sqrt2} &\frac{1}{\sqrt2}
\end{array} \right) \left( \begin{array}{cc}
M_R & M_D \\
M_D & M_L
\end{array} \right) \left( \begin{array}{cc}
\frac{1}{\sqrt2} & - \frac{1}{\sqrt2} \\
\frac{1}{\sqrt2} & \frac{1}{\sqrt2}
\end{array} \right) \nn \\
&= \left( \begin{array}{cc}
M_D + \frac{M_L + M_R}{2} & \frac{M_L - M_R}{2} \\
\frac{M_L - M_R}{2} & - M_D + \frac{M_L + M_R}{2}
\end{array} \right) \equiv \left( \begin{array}{cc}
M_D + m_+ & m_- \\
m_- & - M_D + m_+
\end{array} \right),
\label{eq:cgd-mtrx_tree}
\end{align}
where $m_- \lesssim m_+ \ll M_D$. 
$M'_N$ is then block diagonalized as follows:
\begin{align}
&(\Omega P)^T M'_N (\Omega P) = \left( \begin{array}{cc}
M_D - m_+ & {\bf 0} \\
{\bf 0} & M_D + m_+
\end{array} \right) \equiv M''_N, \\
&\Omega P U_N = \left( \begin{array}{cc}
{\bf 1} & - \frac{1}{2} M^{-1}_D m_- \\
\frac{1}{2} m_-^{\dag} M^{-1}_D & {\bf 1}
\end{array} \right) \left( \begin{array}{cc}
{\bf 0} & {\bf 1} \\
i \cdot {\bf 1} & {\bf 0}
\end{array} \right).
\end{align}
Here, bold number ${\bf 0} ({\bf 1})$ represents $3 \times 3$ zero (unit) matrix. 
Finally, the above mass matrix is diagonalized by
\begin{align}
U_N^T M''_N U_N = \left( \begin{array}{cc}
U_R^T & {\bf 0} \\
{\bf 0} & U_L^T
\end{array} \right) \left( \begin{array}{cc}
M_D - m_+ & {\bf 0} \\
{\bf 0} & M_D + m_+
\end{array} \right) \left( \begin{array}{cc}
U_R & {\bf 0} \\
{\bf 0} & U_L
\end{array} \right) = \left( \begin{array}{cc}
D_R & {\bf 0} \\
{\bf 0} & D_L^C
\end{array} \right).
\end{align}
$M_N$ is approximately diagonalized via $V_N \equiv V^{(1)}_N \Omega P U_N$ as $V_M^T M_N V_N \sim {\rm diag} (D_R, D_L^C)$. 
Note that we have used an assumption that $M_D$ should be proportional to the $3 \times 3$ unit matrix in order to achieve our diagonalization~\cite{Catano:2012kw, Kajiyama:2012xg};
\begin{align}
M_D \approx M_0 \times {\bf 1}.
\label{eq:MDsimdiag}
\end{align}
Then, $M_D \pm m_+$ is given by
\begin{align}
M_D \pm m_+ \simeq \left( \begin{array}{ccc} 
M_0 \pm \delta_{11} & \pm \delta_{12}& \pm \delta_{13} \\
\pm \delta_{12} & M_0 \pm \delta_{22} & \pm \delta_{23} \\
\pm \delta_{13} & \pm \delta_{23} & M_0 \pm \delta_{33} \\
\end{array} \right).
\end{align}
Since the diagonal elements are greater than all the off-diagonal elements, $M_D\pm m_+$ is approximately diagonalized by the following mixing matrices:
\begin{align}
U_{R(L)} &\approx \left( \begin{array}{ccc} 
1 & \epsilon_{12}^{R(L)} & \epsilon_{13}^{R(L)} \\
- \epsilon_{12}^{R(L)} & 1 & \epsilon_{23}^{R(L)}\\
- \epsilon_{13}^{R(L)} & - \epsilon_{23}^{R(L)}& 1
\end{array} \right), \quad \epsilon_{ij}^{R(L)} \approx - \frac{\delta_{ij}}{\delta_{ii} - \delta_{jj}}, \quad (i \neq j) \ .
\end{align}
Since $\epsilon^{R(L)}$ is expected to be tiny, we suppose $|\delta_{ij}| \ll |\delta_{ii}|$ for $i \neq j$ with $i/j = 1, 2, 3$. 
As a result, we find mass eigenstates for $N_{R, L}$ as
\begin{align}
N_{R_a} \equiv \sum_{\alpha = 1} ^6 (V_N)_{a \alpha} \psi_{R_{\alpha}}, \quad N^C_{L_a} \equiv \sum_{\alpha = 1}^6 (V_N)_{a + 3, \alpha} \psi_{R_{\alpha}}.
\label{eq:masseigenNRL}
\end{align}
Hereafter, mass eigenvalues for $\psi_{R_{\alpha}}$ are indicated by $D_{\alpha}$.

\subsection{Charged-lepton mass matrix}

The mass matrix of charged-leptons is induced at one-loop level via the following relevant Lagrangian:
\[
f_{ia}^R \overline{L_{L_i}} \tilde{\eta} N_{R_a} + g_{bj}^L \overline{N_{L_b}} \ell_{R_j} S^+ + f_{ia}^L \overline{L_{L_i}} \tilde{\eta} N_{L_a}^C + g_{bj}^R \overline{N_{R_b}^C} \ell_{R_j} S^+ + {\rm H.c.},
\]
where $\mu_0 H^T (i \tau_2) \eta S^- $ plays a role in connecting $S^{\pm}$ and $\eta^{\pm}$ to construct the one-loop diagram.\footnote{Hereafter, we compute the loop diagrams with a method of mass insertion approximation. 
In particular, we do not perform diagonalizatoin for new scalar particles.} 
Then, these terms are rewritten in terms of mass eigenstates of $N_{L,R}$ [$\psi_{R_{\alpha}}$ in Eq.~\eqref{eq:masseigenNRL}] as follows (summation for $a, b$ are implicitly understood):
\begin{align}
&\frac{f_{ia}^R (V_N)_{a \alpha}}{\sqrt{2}} \overline{\ell_{L_i}} \eta^- \psi_{R_{\alpha}} + (V_N^T)_{\alpha, b + 3} g_{bj}^L \overline{\psi^C_{R_{\alpha}}} \ell_{R_j} S^+ + \frac{f_{ia}^L (V_N)_{a + 3, \alpha}}{\sqrt{2}} \overline{\ell_{L_i}} \eta^- \psi_{R_{\alpha}} + (V_N^T)_{\alpha, b} g_{bj}^R \overline{\psi^C_{R_{\alpha}}} \ell_{R_j} S^+ \nn \\
&\equiv F_{i \alpha} \overline{\ell_{L_i}} \eta^- \psi_{R_{\alpha}} + G_{\alpha j} \overline{\psi^C_{R_{\alpha}}} \ell_{R_j} S^+,
\end{align}
where we define $F_{i \alpha} \equiv \sum_{a = 1}^3 \left[ f_{ia}^R (V_N)_{a \alpha} + f_{ia}^L (V_N)_{a + 3, \alpha} \right] / \sqrt{2}$ and $G_{\alpha j} \equiv \sum_{b = 1}^3 \left[ (V_N^T)_{\alpha, b + 3} g_{bj}^L + (V_N^T)_{\alpha, b} g_{bj}^R \right]$. 
The resultant mass matrix for charged-lepton shown in Fig.~\ref{fig:ClepMass} is given by
\begin{align}
(m_{\ell})_{ij} &= \frac{v_H \mu'_0}{\sqrt{2} (4 \pi)^2 M_0} \frac{ \tilde{F}_{i \alpha} \tilde{D}_{\alpha} \tilde{G}_{\alpha j}}{\tilde{D}_{\alpha}^2 - \tilde{m}_S^2} \left[ \frac{\tilde{D}_{\alpha}^2}{\tilde{D}_{\alpha}^2 - \tilde{m}_S^2} \ln \left( \frac{\tilde{D}^2_{\alpha}}{\tilde{m}_S^2} \right) + \frac{\tilde{D}_{\alpha}^2}{\tilde{D}_{\alpha}^2 - \tilde{m}_{\eta}^2} \ln \left( \frac{\tilde{D}_{\alpha}^2}{\tilde{m}_{\eta}^2} \right) \right],
\end{align}
where $\mu'_0 \equiv f^R_{33} g^R_{33} \mu_0$, $M_0$ is a typical scale of $M_D$ [see Eq.~\eqref{eq:MDsimdiag}], $\tilde{m}_{S, \eta} \equiv m_{S, \eta} / M_0$, $\tilde{D}_{\alpha} \equiv D_{\alpha} / M_0$, $\tilde{F} \equiv F / f^R_{33}$, and $\tilde{G} \equiv G / g^R_{33}$ with $f^R_{33}, \ g^R_{33}$ being, respectively, $(3, 3)$ component of $f^R$ and $g^R$. 
\begin{figure}
\centering
\begin{tikzpicture}
\begin{feynman}
\vertex[label=left:\(L_L \, (b)\)] (a) at (0,0);
\vertex[label=right:\(\ell_R \, (a)\)] (b) at (6,0);
\vertex (c) at (1,0);
\vertex[label=above:\(\psi_R \, (a)\)] (m) at (3,0);
\vertex (d) at (5,0);
\vertex (e) at (3,2);
\vertex[label=above:\(\langle H \rangle\)] (f) at (3,2.7);
\diagram* {
(a) -- (c) -- [insertion={[size=2.5pt]0.5}] (d) -- (b),
(c) -- [scalar, quarter left, edge label=\(\eta \, (b)\)] (e) -- [scalar, quarter left, edge label=\(S^- \, (b)\)] (d),
(f) -- [scalar] (e),
};
\end{feynman}
\end{tikzpicture}
\caption{Charged-lepton masses at one-loop level. 
$a, b$ in the parentheses are the ${\bf Z_5^{NI}}$ assignment for corresponding fields, $\langle H \rangle$ denotes the VEV of the SM Higgs doublet, and we omit the generation indices for fermions. 
It is clear that these effective masses have $a \otimes b$ which does not have $\mathbbm{I}$, and therefore, our fusion rule is radiatively broken by these charged-lepton masses. }
\label{fig:ClepMass}
\end{figure}
We define the diagonalization of $m_{\ell}$ as $D_{\ell} \equiv V^{\dag}_{e_L} m_{\ell} V_{e_R}$.

\subsection{Active neutrino mass matrix}

The active neutrino mass is induced at one-loop level through interaction terms with couplings $F$ and $(H^{\dag} \eta)^2$, and given by~\cite{Ma:2006km}
\begin{align}
(m_{\nu})_{ij} &= \frac{(f^R_{33})^2 M_0}{(4\pi)^2} \tilde{F}_{ia} \tilde{D}_{\alpha} \tilde{F}^T_{aj} \left[ \frac{\tilde{m}_R^2}{\tilde{m}_R^2 - \tilde{D}_{\alpha}^2} \ln \left[ \frac{\tilde{m}_R^2}{\tilde{D}_{\alpha}^2} \right] - \frac{\tilde{m}_I^2}{\tilde{m}_I^2 - \tilde{D}_{\alpha}^2} \ln \left[ \frac{\tilde{m}_I^2}{\tilde{D}_{\alpha}^2} \right] \right] \\
&\equiv \kappa_{\nu} (\tilde{m}_{\nu})_{ij}, 
\label{eq:neutmass}
\end{align}
where $\kappa_{\nu} \equiv (f^R_{33})^2 M_0$, $m_R \equiv M_0 \tilde{m}_R$ is the mass of $\eta_R$ and $m_I \equiv M_0 \tilde{m}_I$ is that of $\eta_I$. 
Corresponding diagram is shown in Fig.~\ref{fig:NuMass}. 
\begin{figure}
\centering
\begin{tikzpicture}
\begin{feynman}
\vertex[label=left:\(L_L \, (b)\)] (a) at (0,0);
\vertex[label=right:\(L_L \, (b)\)] (b) at (6,0);
\vertex (c) at (1,0);
\vertex[label=above:\(\psi_R \, (a)\)] (m) at (3,0);
\vertex (d) at (5,0);
\vertex (e) at (3,2);
\vertex[label=above:\(\langle H \rangle\)] (f) at (1.5,3.2);
\vertex[label=above:\(\langle H \rangle\)] (g) at (4.5,3.2);
\diagram* {
(a) -- (c) -- [insertion={[size=2.5pt]0.5}] (d) -- (b),
(c) -- [scalar, quarter left, edge label=\(\eta \, (b)\)] (e) -- [scalar, quarter left, edge label=\(\eta \, (b)\)] (d),
(f) -- [scalar] (e) -- [scalar] (g),
};
\end{feynman}
\end{tikzpicture}
\caption{Neutrino masses at the one-loop level, with the same notation as in Fig.~\ref{fig:ClepMass}. 
It is clear that these effective masses have $b \otimes b$ which include $\mathbbm{I}$, and therefore, these do not break our fusion rule. }
\label{fig:NuMass}
\end{figure}
We define
$m_{\nu}$ is diagonalized by $D_{\nu} = U^T_{\nu} m_{\nu} U_{\nu}$. 
Then, the observed mixing matrix $U \equiv V_{e_L}^{\dag} U_{\nu}$ and 
$\kappa_{\nu}$ can be rewritten in terms of rescaled neutrino mass eigenvalues $\tilde{D}_{\nu} (\equiv D_{\nu} / \kappa_{\nu})$ and atmospheric neutrino mass-squared difference $\Delta m_{\rm atm}^2$ as follows:
\begin{align}
({\rm NH}):\ \kappa_{\nu}^2 = \frac{|\Delta m_{\rm atm}^2|}{\tilde{D}_{\nu_3}^2 - \tilde{D}_{\nu_1}^2}, \quad ({\rm IH}):\ \kappa_{\nu}^2 = \frac{|\Delta m_{\rm atm}^2|}{\tilde{D}_{\nu_2}^2 - \tilde{D}_{\nu_3}^2},
\end{align}
where NH and IH, respectively, stand for the normal and inverted hierarchies, respectively. 
Subsequently, the solar neutrino mass-squared difference is determined by the relation
\begin{align}
\Delta m_{\rm sol}^2 = \kappa_{\nu}^2 ({\tilde{D}_{\nu_2}^2 - \tilde{D}_{\nu_1}^2}).
\end{align}
The effective mass for neutrinoless double beta decay can be written by 
\begin{align}
\langle m_{ee} \rangle = \kappa_{\nu} |\tilde{D}_{\nu_1} \cos^2 \theta_{12} \cos^2 \theta_{13} + \tilde{D}_{\nu_2} \sin^2 \theta_{12} \cos^2 \theta_{13} e^{i \alpha_{2}} + \tilde{D}_{\nu_3} \sin^2 \theta_{13} e^{i (\alpha_{3} - 2 \delta_{CP})}|,
\end{align}
where $\theta_{12, 23, 13}$ are mixing angles for $U$ in the standard parametrization, $\delta_{CP}$ is a Dirac $CP$ phase, and $\alpha_{2, 3}$ are Majorana phases, defined in Majonara phase matrix $P$ as $P \equiv {\rm diag} (1, e^{\alpha_2 / 2}, e^{\alpha_3 / 2})$. 
A predicted value of $\langle m_{ee} \rangle$ is constrained by the current KamLAND-Zen data; $\langle m_{ee} \rangle < (28-122) \, {\rm meV}$ at 90 \% confidence level~\cite{KamLAND-Zen:2024eml}. 
Furthermore the sum of neutrino masses is constrained by the minimal standard cosmological model $\Lambda_{\rm CDM} + \sum D_{\nu}$ that provides the upper bound $\sum D_{\nu} \le 120 \, {\rm meV}$~\cite{Vagnozzi:2017ovm, Planck:2018vyg}, although it becomes weaker if the data are analyzed in the context of extended cosmological models~\cite{ParticleDataGroup:2014cgo}. 
Recently, DESI and CMB data combination provides more stringent upper bound on the sum as $\sum D_{\nu} \le 72 \, {\rm meV}$~\cite{DESI:2024mwx}. 
Direct search for neutrino mass is done by the Karlsruhe Tritium Neutrino (KATRIN) experiment~\cite{KATRIN:2021uub}, which is the first sub-eV sensitivity on $m_{\nu_e}^2 = (0.26 \pm 0.34) \, {\rm eV}^2$ at 90 \% CL. 
Here, $m_{\nu_e}^2$ is defined by
\begin{align}
\kappa_{\nu}^2 \left[ (\tilde{D}_{\nu_1} \cos \theta_{13} \cos \theta_{12})^2 + (\tilde{D}_{\nu_2} \cos \theta_{13} \sin \theta_{12})^2 + (\tilde{D}_{\nu_3} \sin \theta_{13})^2 \right].
\end{align}
In numerical analysis, we will adopt NuFit 6.0~\cite{Esteban:2024eli} to experimental ranges of neutrino observables in addition to the above constraints of neutrino masses.

\subsection{LFVs, charged-lepton $g-2$ and EDMs}
\label{sec:pheno}

In this subsection, we will discuss LFVs, charged-lepton $g-2$ and EDMs, in order to compare with the experimental constraints in the numerical analyses. 
These contributions are arisen from the one-loop diagram adding external photon to charged scalar line in the diagram of Fig.~\ref{fig:ClepMass}. 
Therefore, these contributions also violate our fusion rule of ${\bf Z_5^{NI}}$. 
The dominant LFVs and charged-lepton $g-2$ and EDMs are induced via the following terms:
\begin{align}
F'_{i \alpha} \overline{\ell_{L_i}} \psi_{R_{\alpha}} \eta^- + G'_{\alpha j} \overline{\psi^C_{R_{\alpha}}} \ell_{R_j} S^+ + {\rm H.c.},
\end{align}
where $F' \equiv V^{\dag}_{e_L} F$, $G' \equiv G V_{e_R}$, and all the above particles are supposed to be mass eigenstates. 
Then, the branching ratios for LFVs, charged-lepton $g-2$ and EDMs are, respectively, given by
\begin{align}
&{\rm BR} (\ell_i \to \ell_j \gamma) \approx 
\frac{48 \pi^3 C_{ij} \alpha_{\rm em}}{G^2_F m_{\ell_i}^2} \left[ |a_{R_{ij}}|^2 + |a_{L_{ij}}|^2 \right], \\
&\Delta a_{\ell_i} \approx - m_{\ell_i}
\left[ a_{R_{ii}} + a_{L_{ii}} \right], \quad d_{\ell_i} = \frac{e}{2} {\rm Im} \left[ a_{R_{ii}} - a_{L_{ii}} \right],
\end{align}
where $G_F$ is Fermi constant, $\alpha_{\rm em} \approx 1 / 137$ is fine structure constant, and $C_{21} \approx 1$, $C_{31} \approx 0.1784$, $C_{32} \approx 0.1736$, $(\ell_1, \ell_2, \ell_3) \equiv (e, \mu, \tau)$. 
$a_R$ and $a_L$ in our model are, respectively, computed as follows:
\begin{align}
a_{R_{ij}} &\approx \frac{\sqrt{2} v_H \mu'_0}{(4 \pi)^2 M^3_0} \tilde{F}'_{j \alpha} \tilde{D}_{\alpha} \tilde{G}'_{\alpha i} F_{lfvs}^{\alpha}, \quad a_{L_{ij}} \approx \frac{\sqrt{2} v_H \mu'_0}{(4 \pi)^2 M^3_0} \tilde{G}'^{\dag}_{j \alpha} \tilde{D}_{\alpha} \tilde{F}'^{\dag}_{\alpha i} F_{lfvs}^{\alpha}, \label{eq:aRaL} \\[0.5ex]
F_{lfvs}^{\alpha} &= \int dx_1 dx_2 dx_3 dx_4 x_5 \delta(1-x_1-x_2-x_3-x_4-x_5) \left( \frac{3}{\Delta_{\alpha}^2} + \frac{\tilde{m}_S^2 + \tilde{m}_{\eta}^2}{\Delta_{\alpha}^3} \right),
\end{align}
where $\Delta_{\alpha} \equiv x_1 \tilde{D}_{\alpha}^2 + (x_2 + x_4) \tilde{m}_S^2 + (x_3 + x_5) \tilde{m}_{\eta}^2$ and $x_i$s are Feynman parameters. 

Experimental upper bounds and future prospects for the branching ratios of the LFV decays are given by~\cite{MEG:2016leq, BaBar:2009hkt,Renga:2018fpd,MEGII:2023ltw}
\begin{align}
&{\rm Current}: {\rm BR} (\mu \to e \gamma) \lesssim 3.1 \times 10^{-13}, \quad {\rm BR} (\tau \to e \gamma) \lesssim 3.3 \times 10^{-8}, \quad {\rm BR} (\tau \to \mu \gamma) \lesssim 4.4 \times 10^{-8}, \\
&{\rm Future}: {\rm BR} (\mu \to e \gamma) \lesssim 6.0 \times 10^{-14}, \quad {\rm BR} (\tau \to e \gamma) \lesssim 9.0 \times 10^{-9}, \quad {\rm BR} (\tau \to \mu \gamma) \lesssim 6.9 \times 10^{-9}.
\end{align}
Here, we estimate our input parameters to satisfy the LFVs to prepare the range of numerical analysis in the next section, focusing on the most stringent bound on $\mu \to e \gamma$ for safety. 
When we simply fix the following values $\tilde{F}' = \tilde{G}' = 0.01$, $\tilde{D}_{\alpha} F^{\alpha}_{lfvs} = 1$, $\mu'_0 = 1 \, {\rm GeV}$ and $M_0 = 1000 \, {\rm GeV}$, ${\rm BR} (\mu \to e \gamma) \approx 6.95 \times 10^{-13}$ that is larger by factor of $\approx 2.24$ than the experimental upper bound. 
However, as one can understand from Eq.~\eqref{eq:aRaL}, the branching ratio hardly depends on the value of $M_0$, ${\rm BR} (\mu \to e \gamma) \propto 1 / M_0^6$. 
Using this fact, ${\rm BR} (\mu \to e \gamma) \lesssim 3.1 \times 10^{-13}$ can be easily obtained by setting $M_0 \gtrsim 1144 \, {\rm GeV}$, and thus, it is easy to satisfy the LFVs as well as other bounds. 
Experimental results for charged-lepton $g-2$ at 1$\sigma$ are given by~\cite{Fan:2022oyb,Fan:2022eto,ParticleDataGroup:2024cfk,Muong-2:2021ojo,Muong-2:2023cdq,Aliberti:2025beg,Muong-2:2025xyk,Morel:2020dww}\footnote{For $\Delta a_e$, there is another result by using the fine-structure constant $\alpha$ determined by ${}^{133}{\rm C_s}$, whose result has opposite sign, $\Delta a_e = (-10.1 \pm 2.7) \times 10^{-13}$~\cite{Parker:2018vye}.}
\begin{align}
& {\rm Current}: \Delta a_e \simeq (3.41 \pm 1.64) \times 10^{-13}, \quad \Delta a_{\mu} \simeq (38 \pm 63) \times 10^{-11}, \\
& {\rm Past}: \Delta a_e \simeq (4.8 \pm 3.0) \times 10^{-13}, \quad \Delta a_{\mu} \simeq (25.1 \pm 5.9) \times 10^{-10} \ {\rm or} \ (24.9 \pm 4.8) \times 10^{-10}.
\end{align}
Experimental upper bounds for EDMs are, respectively, given by~\cite{Abada:2024hpb}
\begin{align}
&|d_e| < 4.1 \times 10^{-30} \ \ecm \ ({\rm JILA}), \quad |d_{\mu}| < 1.9 \times 10^{-19} \ \ecm \ ({\rm Muon\ g-2}), \\
&|{\rm Re} (d_{\tau})| < 4.5\times 10^{-17} \ \ecm \ ({\rm Belle}), \quad |{\rm Im} (d_{\tau})| < 2.5\times 10^{-17} \ \ecm \ ({\rm Belle}).
\end{align}
Furthermore, indirect bounds for the muon and tau EDMs provide more stringent constraints than the direct experimental ones, which are obtained by evaluating the muon-loop induced light-by-light $CP$ odd amplitude~\cite{Ema:2021jds, Ema:2022wxd}:
\begin{align}
|d_{\mu}| < 6.3\times 10^{-21} \ \ecm,
%|d_{\mu}| < 1.7\times 10^{-20} \ \ecm,
\quad
|d_{\tau}| < 4.1\times 10^{-19} \ \ecm,
%|d_{\tau}| < 1.1\times 10^{-18} \ \ecm.
\end{align}
by using the latest electron EDM bound. 
Note that these bounds are obtained by the current electron EDM bound, and therefore, an improvement of the electron EDM bound will also push down these upper bounds.

\section{Numerical results}
\label{sec:III}

Here, we perform $\Delta \chi^2$ numerical analysis and investigate our allowed regions to satisfy whole the experimental results discussed above. 
Here, we assume all the relevant observables to be the Gaussian distribution under the $\Delta \chi^2$ analysis, and hence,
\begin{align}
\Delta \chi^2 = \sum_{i} \left( \frac{O_i^{\rm obs} - O_i^{\rm th}}{\delta O_i^{\rm exp}} \right)^2, \label{eq:chi-square}
\end{align}
where $O_i^{\rm obs (th)}$ is observed (theoretically obtained) value of corresponding observables and $\delta O_i^{\rm exp}$ indicates the experimental error at $1\sigma$. 
We adopt four reliable observables; $\{ \sin^2 \theta_{12}, \sin^2 \theta_{13}, \Delta m^2_{\rm atm}, \Delta m^2_{\rm sol} \}$, to estimate the $\Delta \chi^2$ analysis and take our allowed regions up to $5\sigma$.\footnote{$3\sigma$ with four degrees of freedom corresponds to $\Delta \chi^2 \approx 34.555$~\cite{Okada:2019lzv}.} 
On the other hand, we independently request $\sin^2 \theta_{23}$ to be within $3\sigma$ level of Nufit 6.0, since the experimental error is deviated from Gaussian distribution. 
In our parameter space, we fix $f^R = f^L$ and $g^R = g^L$ for simplicity.\footnote{Note that, this simplification does not eliminate all physical $CP$ phases in our model: As mentioned in Eq.~\eqref{eq:CPexample}, we still have second example of the physical $CP$ phase, as well as other ones which coming from, e.g., off-diagonal elements of $f^{R (L)}$, $g^{R (L)}$ and $M_{R, L}$. 
Therefore, we can discuss predictions for charged-lepton EDMs in our models.} 
Then, we randomly select our input parameters as follows:
\begin{align}
&10^{-3} \le \frac{|\delta_{11,22,33}|}{\rm GeV} \le 0.1, \quad 10^{-5} \le \frac{|\delta_{12, 13, 23}|}{\rm GeV} \le 10^{-3}, \quad 10^{-5} \le \frac{|m_-|}{\rm GeV} \le 0.1, \label{eq:input1} \\
&10^{-7} \le \delta \tilde{m} \le 10^{-5}, \quad 10^{-5} \le {|f_R|} \le 1, \quad 10^{-5} \le {|g_R|} \le 0.1, \\
&100 \le \frac{M_0}{\rm GeV} \le 10^5, \quad 0.1 \le \frac{\mu'_0}{\rm GeV} \le 10, \quad 0.1 \le \tilde{m}_I \le 10, \quad 10^{-6} \le \tilde m_S \le 10^{-4}, \label{eq:input3}
\end{align}
where $\delta_{ij}$ is the component of $m_+$, $\tilde{m}_R \equiv \tilde{m}_I + \delta \tilde{m}$, $\tilde{m}_{\eta} \equiv \tilde{m}_I$. 
Arguments run over $[- \pi, + \pi]$.\footnote{If there are no symbols of absolute values $|\cdots|$ in Eqs.~\eqref{eq:input1}--\eqref{eq:input3}, such parameters are considered as real parameters.} 
Note that arbitrary three components of $g^R$ can directly be determined by charged-lepton mass matrix putting experimental results for charged-lepton masses. 
Here, we select $g^R_{11, 22, 32}$. 
Starting at the above regions in our input parameters, we specify the parameter space of their absolute values to fit the neutrino mass-square differences, three mixing angles, LFVs, and lepton $g-2$.\footnote{This is because there are too many parameters and too much time to find out all the allowed regions.} 
But, we keep the wide ranges of their arguments that play a crucial role in determining the Dirac $CP$ phase $\delta_{CP}$, two Majorana phases $\alpha_{2, 3}$ and EDMs $d_{e, \mu, \tau}$.

\subsection{NH}
At first, we show our benchmark point that does not depend on phases in Table~\ref{tab:BF_nh}. 
The best fit (BF) value corresponds to $\Delta \chi_{\rm min} \approx 1.06$. 
%%%%%%%%%%%%%%%%%%
\begin{table}[tb]
\setlength\tabcolsep{0.2cm}
\begin{tabular}{c|c||c|c||c|c}
\hline
Parameter & BF & Parameter & BF & Parameter & BF \\ \hline \hline
$s_{12}$ & $0.553$ & $s_{23}$ & $0.763$ & $s_{13}$ & $0.147$ \\ \hline
$\Delta m^2_{\rm sol}$ & $7.32 \times 10^{-23} \, {\rm GeV}^2$ & $\Delta m^2_{\rm atm}$ & $2.54 \times 10^{-21} \, {\rm GeV}^2$ & ~ & ~ \\ \hline
$m_{\nu_e}$ & $27.9 \, {\rm meV}$ & $\sum D_{\nu}$ & $111 \, {\rm meV}$ & $D_{\nu_1}$ & $26.5 \, {\rm meV}$ \\ \hline
$m_e$ & $0.000511 \, {\rm GeV}$ & $m_{\mu}$ & $0.106 \, {\rm GeV}$ & $m_{\tau}$ & $1.777 \, {\rm GeV}$ \\ \hline
$\Delta a_e$ & $8.74 \times 10^{-17}$ & $\Delta a_{\mu}$ & $2.52 \times 10^{-12}$ & ~ & ~ \\ \hline
BR($\mu \to e \gamma$) & $3.54 \times 10^{-24}$ & BR($\tau \to e \gamma$) & $2.35 \times 10^{-27}$ & BR($\tau \to \mu \gamma$) & $4.08 \times 10^{-21}$ \\ \hline
\end{tabular}
\caption{\label{tab:BF_nh}
Best-fit experimental values in the NH case, corresponding to $\Delta \chi_{\rm min} \approx 1.06$. 
Here we display these values that only depend on absolute values. }
\end{table}
%%%%%%%%%%%%%%
It is clear that the branching ratios for LFVs are enough small to satisfy current bounds. 
Moreover, the prediction of $\Delta a_{\mu}$ is within $1\sigma$ of current results, while that of $\Delta a_e$ is within $3\sigma$. 

%%%%%%%%%%%%%%%%%%%
\begin{figure}[tb]
\begin{center}
\includegraphics[width=100mm]{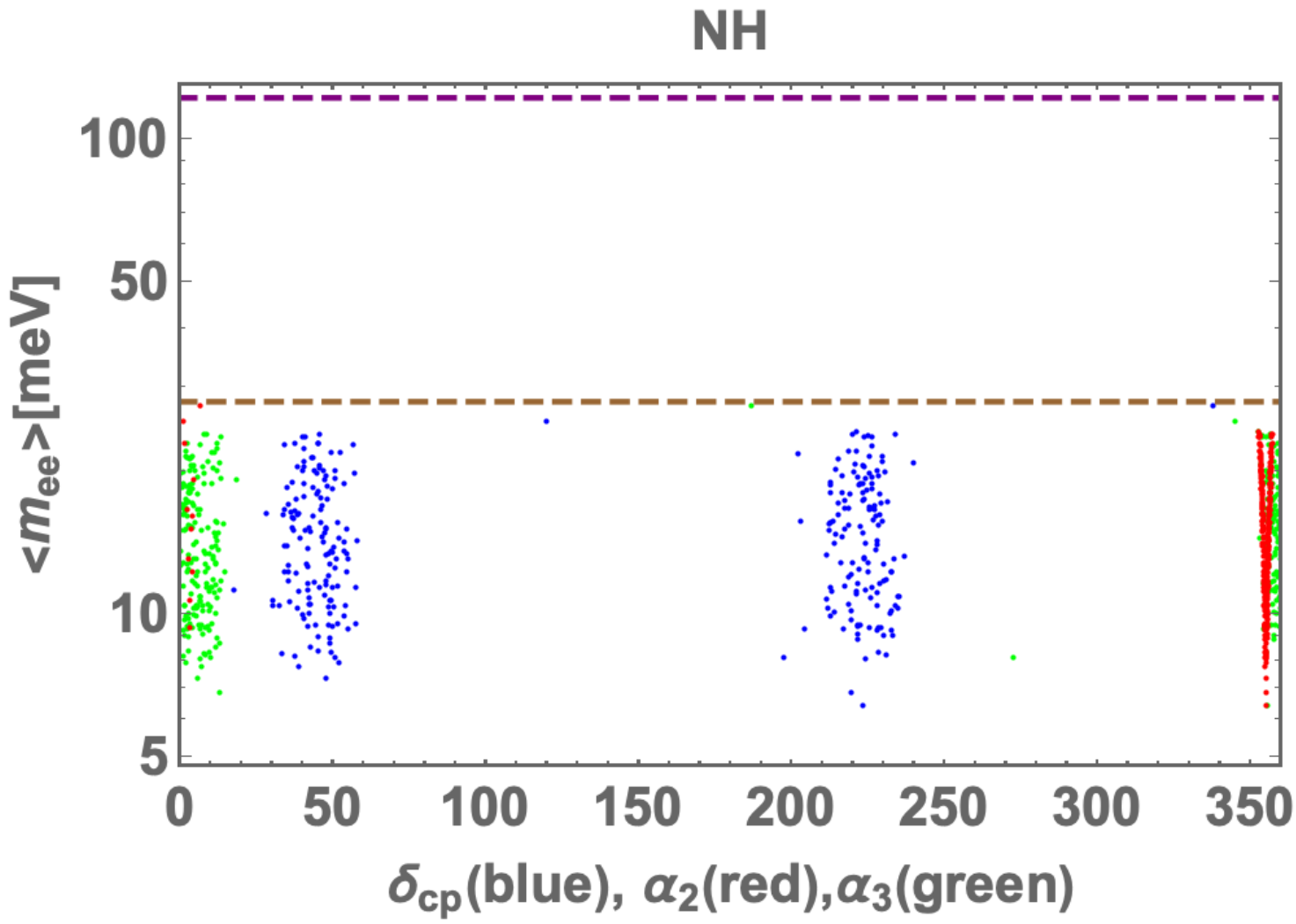}
\caption{Allowed regions for $\langle m_{ee} \rangle$ in terms of phases $\delta_{CP}$ (blue), $\alpha_2$ (red), and $\alpha_3$ (green) in case of NH. 
Brown (lower) and purple (upper) dashed lines correspond to $\langle m_{ee} \rangle = 28$ and $122 \, {\rm meV}$, respectively. } 
\label{fig:nh1}
\end{center}
\end{figure}
%%%%%%%%%%%%%%%%%%%
Under Table~\ref{tab:BF_nh}, we figure out the neutrinoless double beta decay in terms of phases $\delta_{CP}$, $\alpha_2$, and $\alpha_3$ in Fig.~\ref{fig:nh1}, as blue, red and green points, respectively. 
The maximum value of neutrinoless double beta decay just reaches the upper limit of 28 meV.
We find the following allowed region-tendencies of phases; $30^{\circ} \lesssim \delta_{CP} \lesssim 60^{\circ}$, $200^{\circ} \lesssim \delta_{CP} \lesssim 240^{\circ}$, $-10^{\circ} \lesssim \alpha_2 \lesssim 0^{\circ}$ and $-10^{\circ} \lesssim \alpha_3 \lesssim 20^{\circ}$.

%%%%%%%%%%%%%%%%%%%
\begin{figure}[tb]
\begin{center}
\includegraphics[width=56mm]{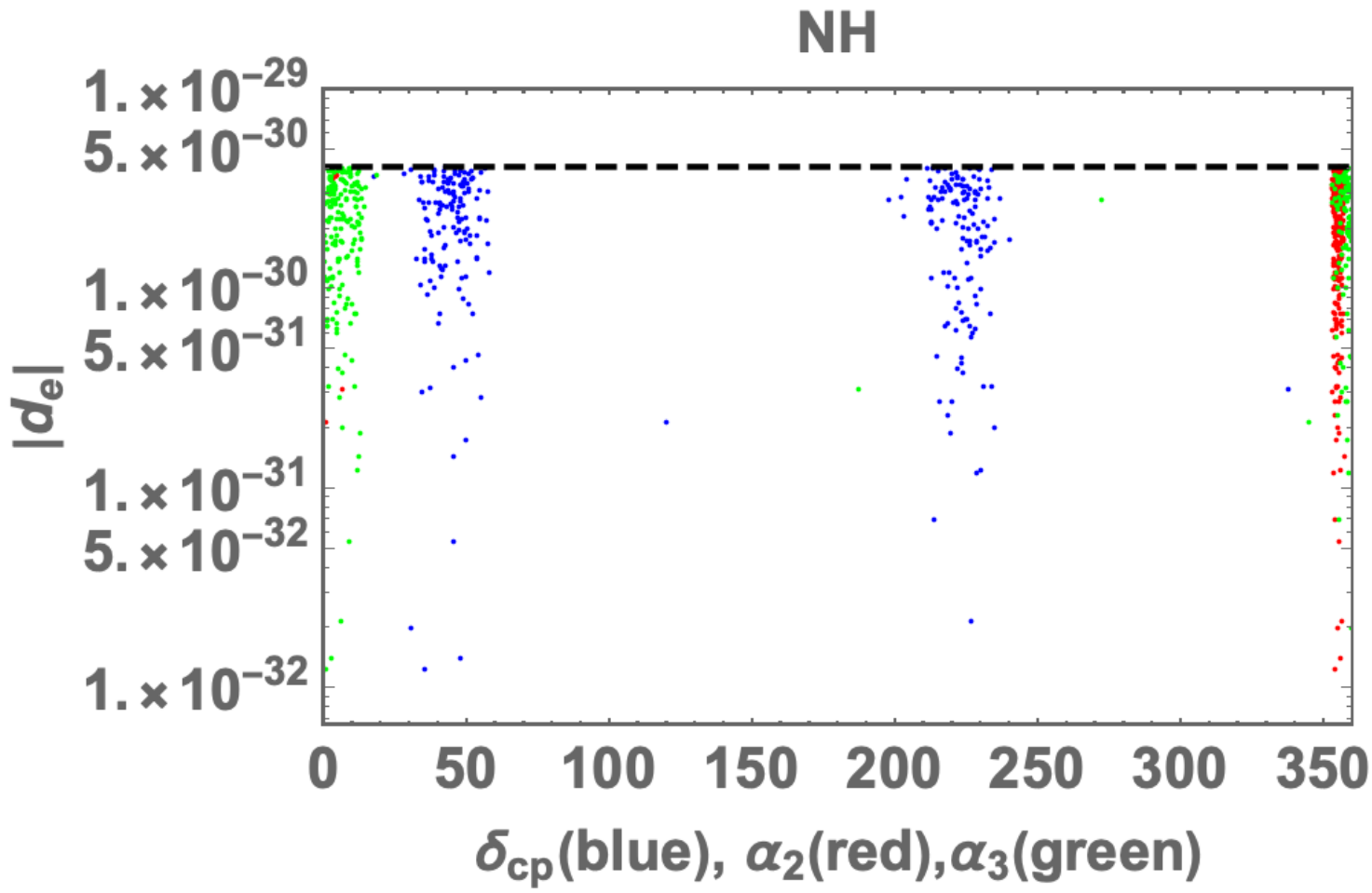}
\includegraphics[width=53mm]{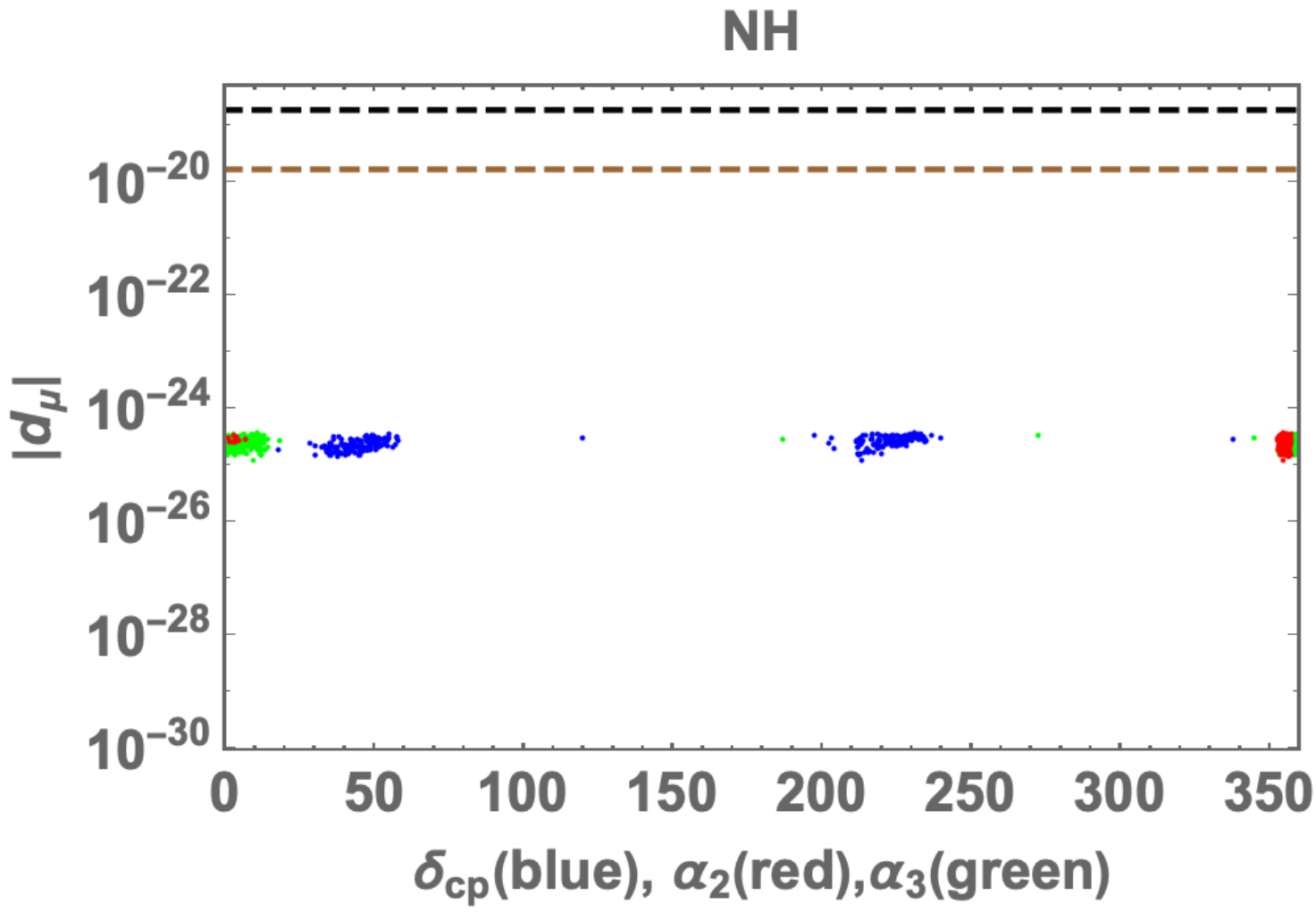}
\includegraphics[width=53mm]{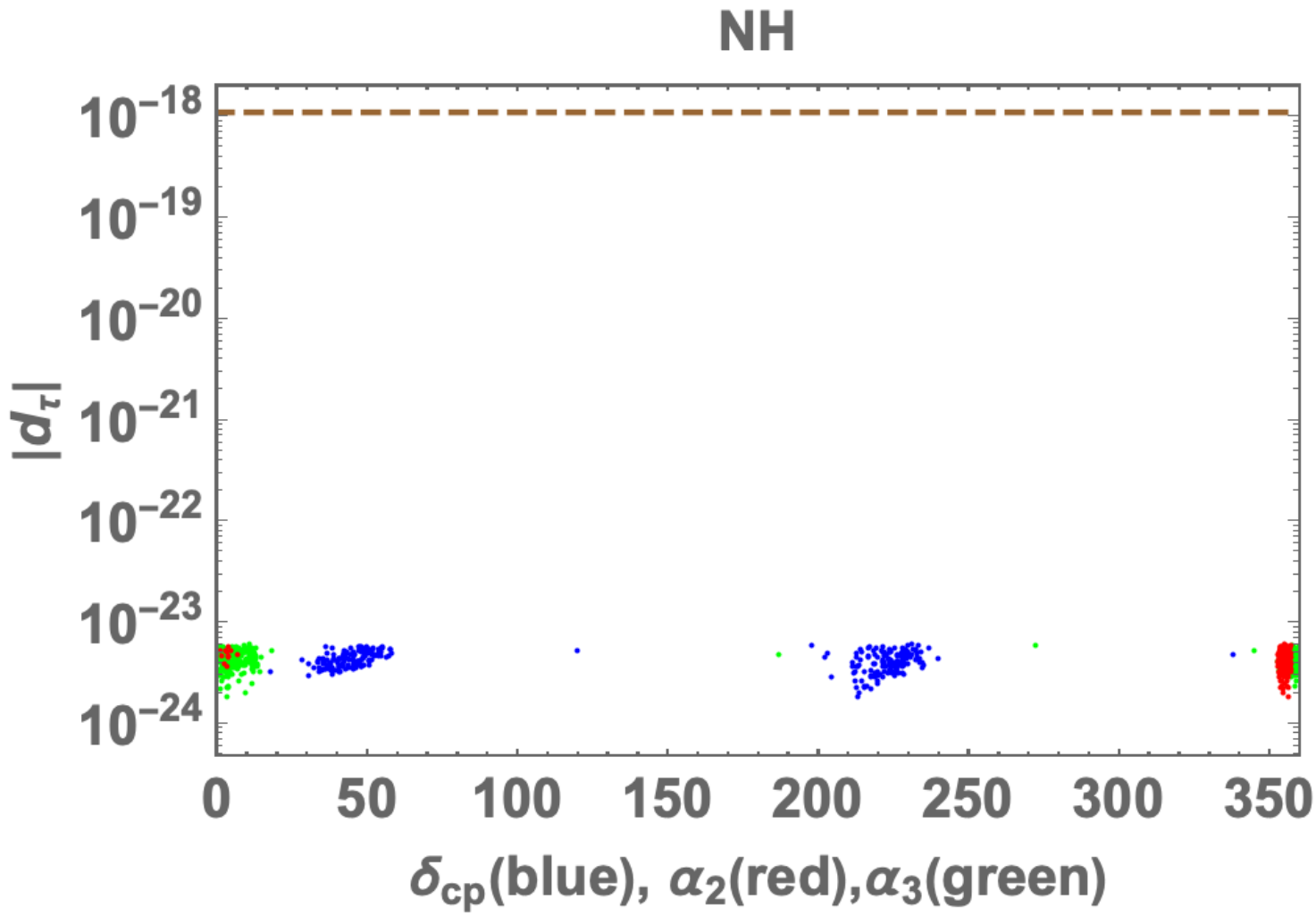}
\caption{Allowed regions for EDMs in terms of phases $\delta_{CP}$, $\alpha_2$, and $\alpha_3$ in case of NH, with same color manner as in Fig.~\ref{fig:nh1}. 
The black dashed line is the direct experimental upper bound, while brown corresponds to the indirect upper bound. } 
\label{fig:nh2}
\end{center}
\end{figure}
%%%%%%%%%%%%%%%%%%%
Next, we show our EDM predictions of $|d_e|$ (left), $|d_{\mu}|$ (center) and $|d_{\tau}|$ (right) for the case of NH in Fig.~\ref{fig:nh2}, in term of three phases, where each of color plots are the same as the one in Fig.~\ref{fig:nh1}. 
These figures suggest that $|d_e|$ reaches the experimental upper bound $4.1 \times 10^{-30} \ \ecm$, while typical values of muon and tau EDM are $|d_{\mu}| \simeq (1$-$4) \times 10^{-25} \ \ecm$ and $|d_{\tau}| \simeq (2$-$6) \times 10^{-24} \ \ecm$. 
Interestingly, the constraint $|d_e| < 4.1 \times 10^{-30} \ \ecm$ results in $\langle m_{ee} \rangle \lesssim 28 \, {\rm meV}$, which is more stringent upper limit of the neutrinoless double beta decay. 
It is also emphasized that the predictions of $|d_{\mu}|$ and $|d_{\tau}|$ are several orders of magnitude larger than those from the minimal-flavor violation relation: $|d_{\mu}| \simeq |d_e| (m_{\mu} / m_e) \lesssim 8.5 \times 10^{-28} \ \ecm$ and $|d_{\tau}| \simeq |d_e| (m_{\tau} / m_e) \lesssim 1.4 \times 10^{-26} \ \ecm$. 
This is because origins of physical $CP$ phases for $d_{e, \mu, \tau}$ are different from each other. 
Another interesting point is mostly fixed predictions of $|d_{\mu}|$ and $|d_{\tau}|$ are obtained by the focused benchmark point. 
This is because they tend to be determined by the arguments whose allowed region is narrow for our input parameters, while $|d_e|$ tends to be predicted by wide ones.\footnote{In fact, we found that some arguments are required to be narrow range for fitting the experimental observables, while some of them are not so strictly constrained. 
Although it might be interesting to explore its dependence, we will not show them in detail, since we have a lot of parameters. 
Note here that absolute values of $d_{\mu}$ and $d_{\tau}$ are much larger than those of $d_e$.} 

\subsection{IH}

Next, we show our benchmark point that does not depend on phases in Table~\ref{tab:BF_ih}. 
The BF value results in $\Delta \chi_{\rm min} \approx 0.539$. 
%%%%%%%%%%%%%%%%%%
\begin{table}[tb]
\setlength\tabcolsep{0.2cm}
\begin{tabular}{c|c||c|c||c|c}
\hline
Parameter & BF & Parameter & BF & Parameter & BF \\ \hline \hline
$s_{12}$ & $0.557$ & $s_{23}$ & $0.695$ & $s_{13}$ & $0.150$ \\ \hline
$\Delta m^2_{\rm sol}$ & $7.60 \times 10^{-23} \, {\rm GeV}^2$ 
& $\Delta m^2_{\rm atm}$ & $2.47 \times 10^{-21} \, {\rm GeV}^2$ & ~ & ~ \\ \hline
$m_{\nu_e}$ & $48.6 \, {\rm meV}$ & $\sum D_{\nu} $ & $98.6 \, {\rm meV}$ & $D_{\nu_3}$ & $0.0166 \, {\rm meV}$ \\ \hline
$m_e$ & $0.000511 \, {\rm GeV}$ & $m_{\mu}$ & $0.106 \, {\rm GeV}$ & $m_{\tau}$ & $1.777 \, {\rm GeV}$ \\ \hline
$\Delta a_e$ & $7.35 \times 10^{-16}$ &$\Delta a_{\mu}$ & $2.95 \times 10^{-11}$ & ~ & ~ \\ \hline
BR($\mu \to e \gamma$) & $3.93 \times 10^{-20}$ & BR($\tau \to e \gamma$) & $1.78 \times 10^{-23}$ & BR($\tau \to \mu \gamma$) & $2.64 \times 10^{-17}$ \\ \hline
\end{tabular}
\caption{\label{tab:BF_ih}
Best-fit experimental values in the IH case, corresponding to $\Delta \chi_{\rm min} \approx 0.539$. 
Here we select the above values that dominantly depend on absolute values. }
\end{table}
%%%%%%%%%%%%%%
Compared with the case of NH, we obtain about one order of magnitude larger $\Delta a_e$ and $\Delta a_{\mu}$. 
On the other hand, we found that the IH case predicts about 4 orders of magnitude larger LFVs than those in the NH case, which still satisfy the current bounds. 

%%%%%%%%%%%%%%%%%%%
\begin{figure}[tb]
\begin{center}
\includegraphics[width=100mm]{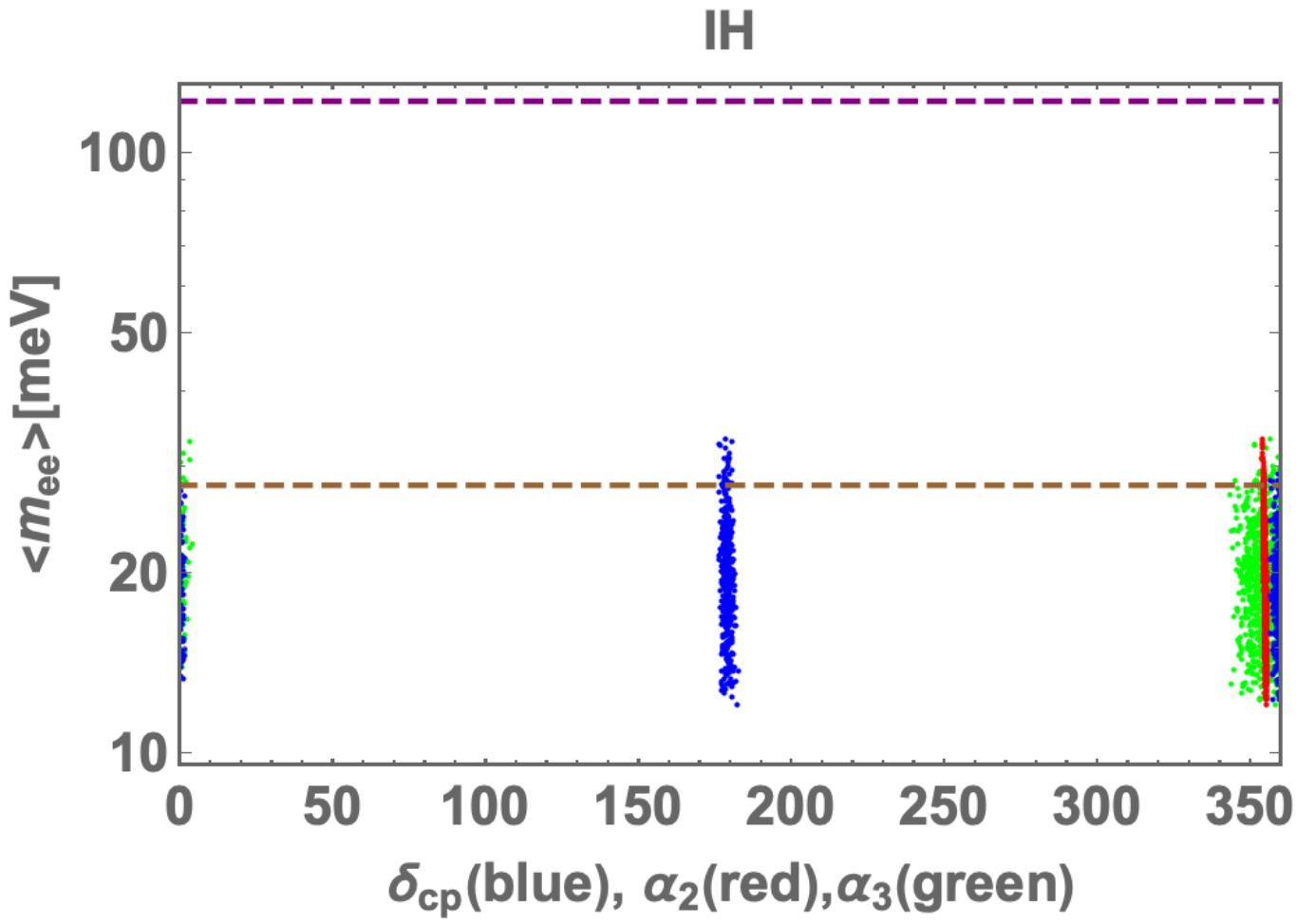}
\caption{Allowed regions for $\langle m_{ee}\rangle$ in terms of phases $\delta_{CP}$ (blue), $\alpha_2$ (red), and $\alpha_3$ (green) in case of IH. 
The meanings of dashed lines are the same as in Fig.~\ref{fig:nh1}. } 
\label{fig:ih1}
\end{center}
\end{figure}
%%%%%%%%%%%%%%%%%%%
Under Table~\ref{tab:BF_ih}, we figure out the neutrinoless double beta decay in terms of phases $\delta_{CP}$, $\alpha_2$, and $\alpha_3$ in Fig.~\ref{fig:ih1}, with same color manner as in Fig.~\ref{fig:nh1}. 
The several points exceed the upper limit of $28 \, {\rm meV}$, and the maximum value reaches $33.5 \, {\rm meV}$.
We find $\delta_{CP}$ is localized at nearby $0^{\circ}$ and $180^{\circ}$, while Majorana phases are $\alpha_2 \sim 355^{\circ}$ and $-20^{\circ} \lesssim \alpha_3 \lesssim 0^{\circ}$. 

%%%%%%%%%%%%%%%%%%%
\begin{figure}[tb]
\begin{center}
\includegraphics[width=53mm]{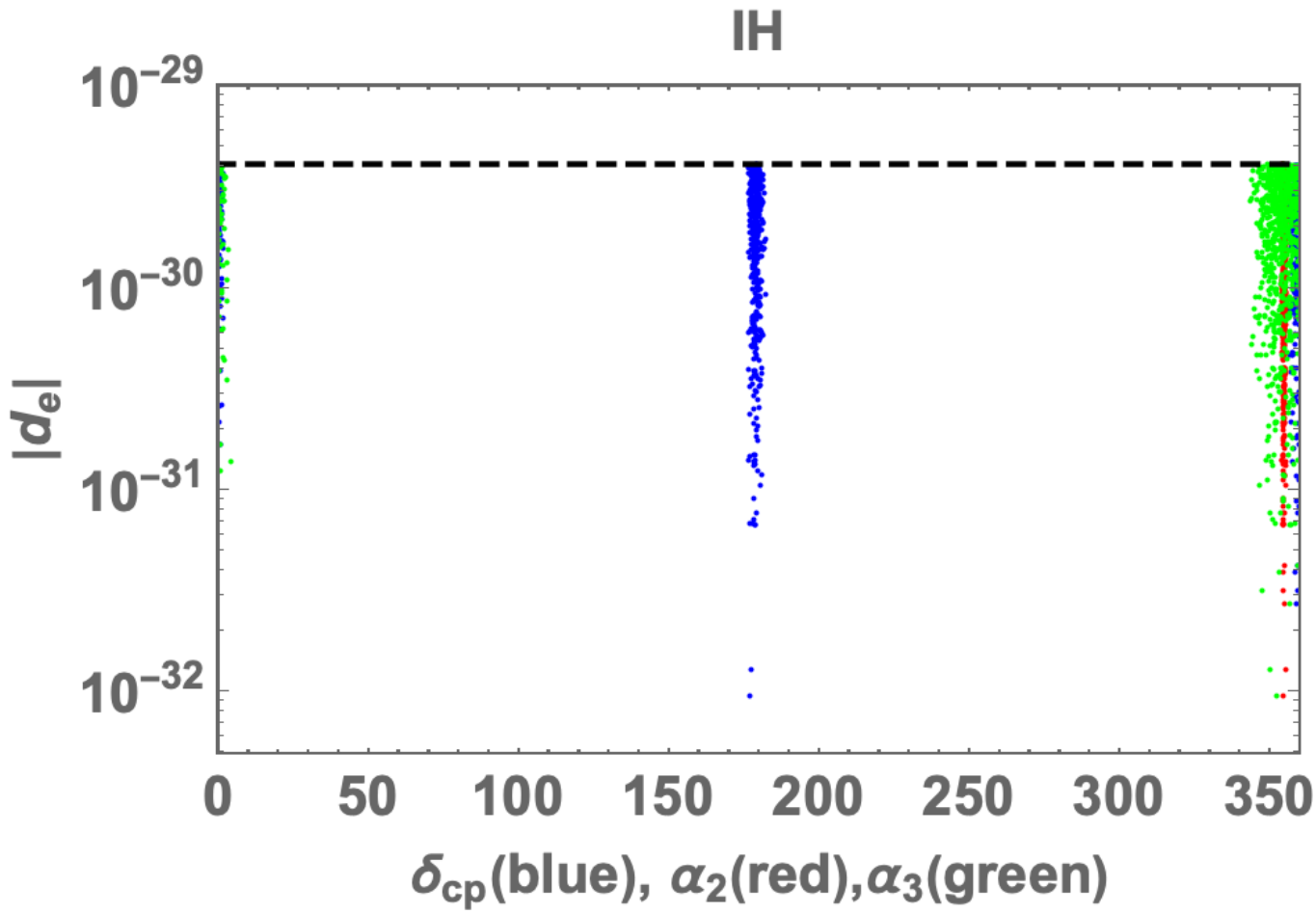}
\includegraphics[width=53mm]{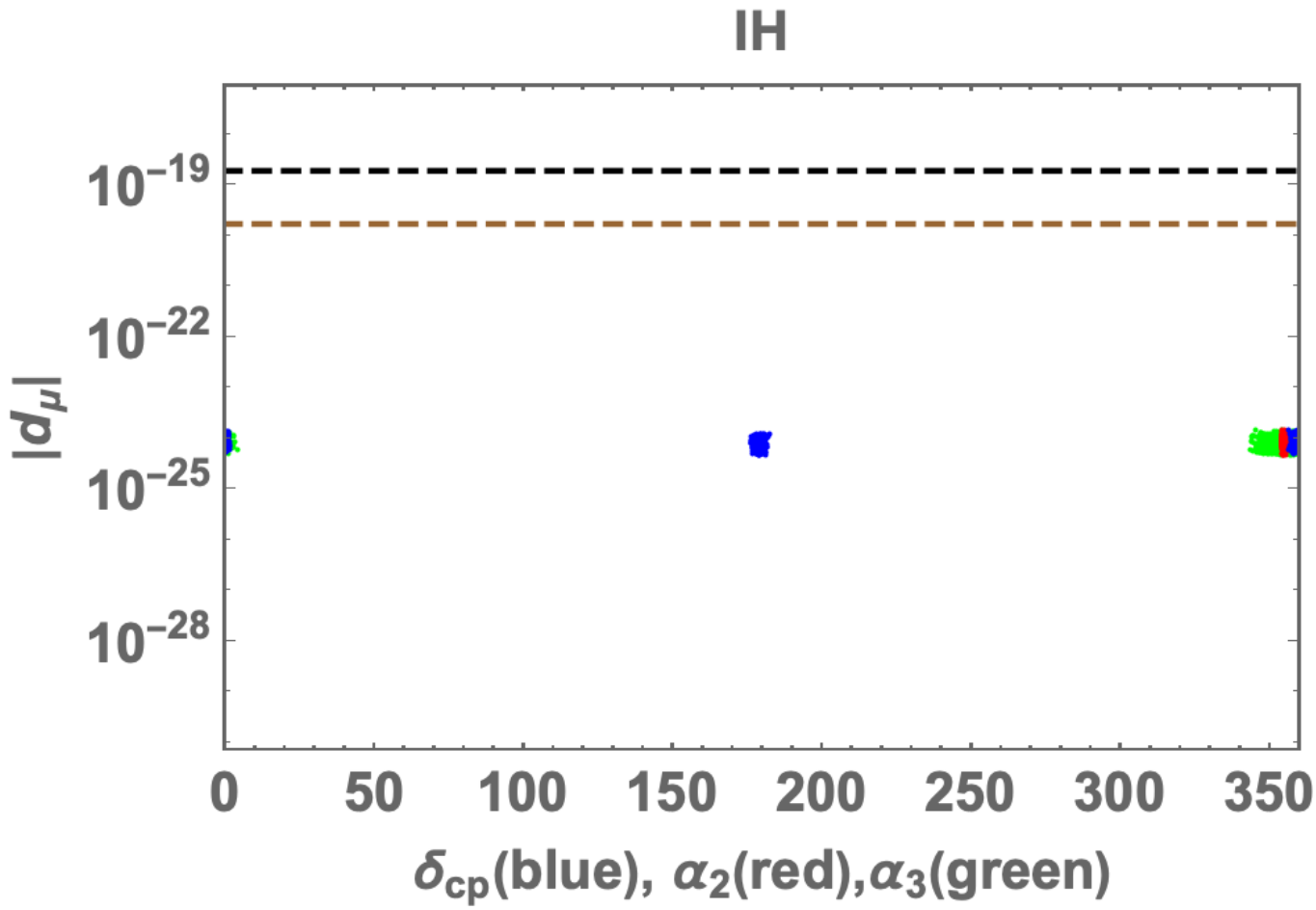}
\includegraphics[width=53mm]{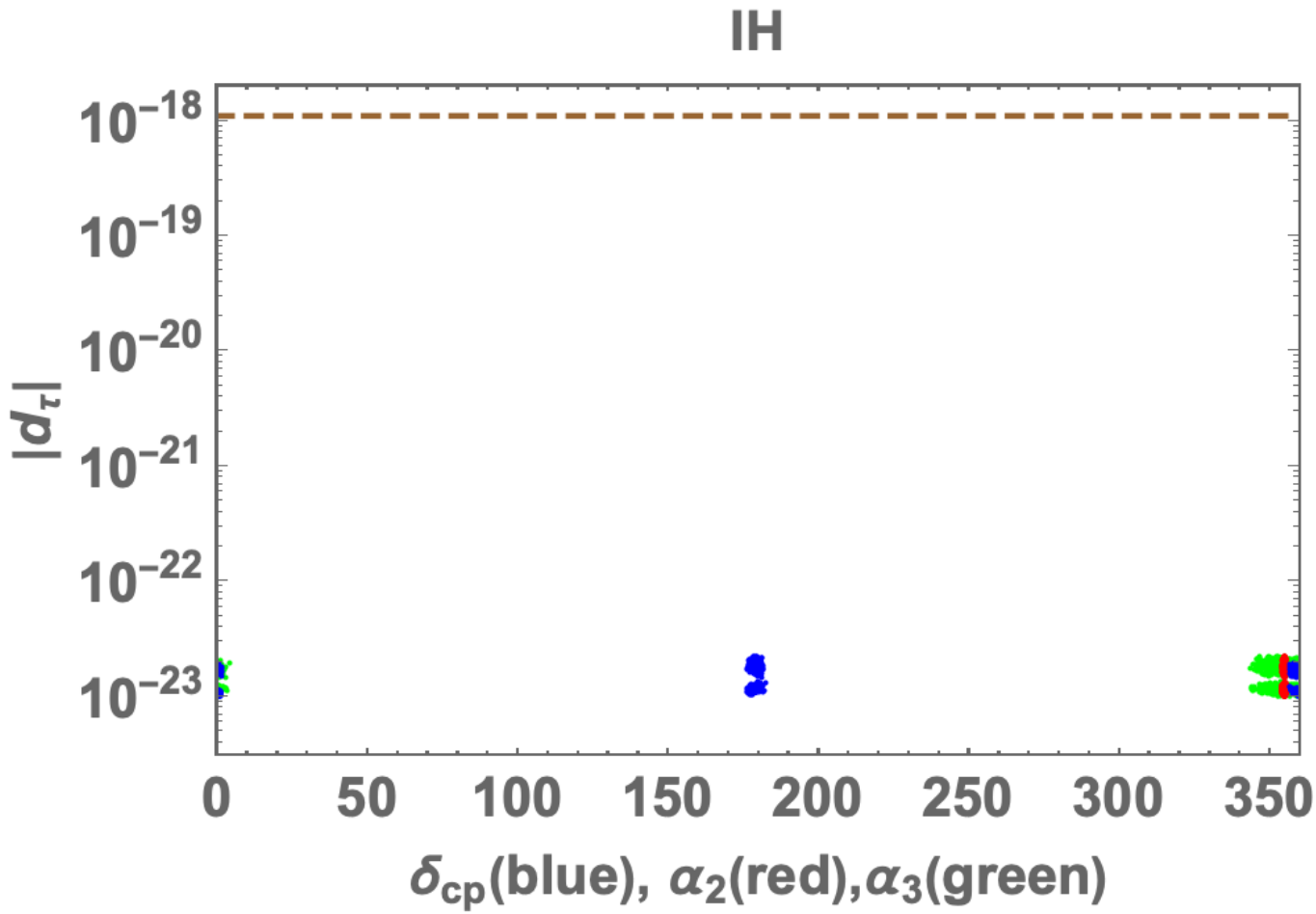}
\caption{Same plot as Fig.~\ref{fig:nh2}, but for the case of IH. }
\label{fig:ih2}
\end{center}
\end{figure}
%%%%%%%%%%%%%%%%%%%
In Fig.~\ref{fig:ih2}, we show EDMs $|d_e|$ (left), $|d_{\mu}|$ (center), and $|d_{\tau}|$ (right), in term of three phases, where each of color plots are the same as the one in Fig.~\ref{fig:nh1}. 
These figures suggest that $|d_e|$ can reach the experimental upper bound $4.1 \times 10^{-30} \ \ecm$, but typical values of $|d_{\mu, \tau}|$ are, respectively, $10^{-24}$ and $10^{-23} \ \ecm$. 
Interestingly, the constraint $|d_e| < 4.1 \times 10^{-30} \ \ecm$ leads us to the similar result of the NH case; $\langle m_{ee} \rangle \lesssim 33.5 \, {\rm meV}$ which is just above the stringent upper limit of the neutrinoless double beta decay. 
The EDM dependencies are similar to those in the NH case.

\section{Summary and discussion}
\label{sec:IV}

We have proposed a radiatively induced lepton mass model by introducing a $Z_2$ gauging $Z_5$ fusion rule to forbid tree-level lepton masses while keeping relevant interactions to induce them via radiative corrections. 
The charged-lepton mass matrix is generated at one loop level via dynamical breaking of the fusion rule. 
On the other hand, the neutrino mass matrix is induced at the one-loop level without breaking the fusion rule. 
These facts lead us to discuss LFVs, electron and muon $g-2$, and charged-lepton EDMs, where these phenomenology originate from the charged-lepton sector. 

Due to too many free parameters, we have specified the absolute values of parameter set in order to fit the neutrino oscillation data and the charged-lepton mass observables. 
In addition, we have fixed the set of best fit values in cases of NH and IH. 
Then, once we have randomly selected the arguments of our free parameters and scanned whole their ranges, we have obtained our observables such as the Dirac $CP$ phase, two Majorana phases, charged-lepton EDMs and the neutrinoless double beta decay, all of which depends on their arguments. 
As results of our numerical analyses, we have found several interesting tendencies; $30^{\circ} \lesssim \delta_{CP} \lesssim 60^{\circ}$, $200^{\circ} \lesssim \delta_{CP} \lesssim 240^{\circ}$, $-10^{\circ} \lesssim \alpha_2 \lesssim 0^{\circ}$ and $-10^{\circ} \lesssim \alpha_3 \lesssim 20^{\circ}$ in NH, and $\delta_{CP} \sim 0^{\circ}, \ 180^{\circ}$, $\alpha_2 \sim 355^{\circ}$ and $-20^{\circ} \lesssim \alpha_3 \lesssim 0^{\circ}$ in IH. 
We have also discovered that the constraint $|d_e| < 4.1 \times 10^{-30} \ \ecm$ has led us to the maximum prediction of $\langle m_{ee} \rangle$ to be $28 \, {\rm meV}$ for NH and $33.5 \ {\rm meV}$ for IH, both of which are around more stringent upper limit of the neutrinoless double beta decay. 
Furthermore, the predictions of the muon and tau EDMs are several orders of magnitude larger than those obtained by the rough minimal-flavor violation relation, due to different origin of the $CP$ phase. 

We would like to mention the application of our mechanism to the quark sector. 
Although we did not focus on the quark sector in this work, we can construct the model for radiative quark masses by assigning the proper noninvertible fusion rule to quarks. 
The simplest model is to assign $b$ to $Q_L$ and $a$ to $d_R$ so that the tree-level down-type Yukawa couplings are forbidden. 
Then, one would obtain the same structure as in the case of our charged-lepton sector. 
Note that the up-type quark sector is not favored to apply this mechanism, since top-Yukawa coupling is order one,\footnote{The radiative masses for up and charm quarks are acceptable, due to small Yukawa couplings.} and its radiative generation will cause the issues, e.g., requiring too large couplings which is not allowed in the sense of perturbativity. 
Therefore, down-type quark masses are generated via one-loop level after radiative breaking of the fusion rule. 
In this case, $b \to s \ell \bar{\ell}$ anomalies can be taken into account through $\eta$, as well as several constraints such as neutral meson mixings and $b \to s \gamma$. 
Detailed analysis is beyond our scope, but it is also worth constructing a radiative quark mass model in future. 

Before closing our paper, we comment on a possible dark matter candidate in the model. 
Potentially, we have two dark matter candidates; the lightest neutral fermion $\psi_{R_1}$ or the lightest neutral component of $\eta$ which are similar to the case of the Ma model. 
However, our allowed mass region through the numerical analysis suggests that these masses are ${\cal O} (1) \, {\rm PeV}$ scale in case of $m_S \sim 100 \, {\rm GeV}$, and DMs decay so quickly that they cannot be DM candidates. 
In case of the bosonic one, it decays into, {\it e.g.}, pairs of $\bar{\ell} \ell$ through $y^{\ell}$. 
In case of the fermionic one, it decays into $\ell^- W^+ \nu$ through $g^{L/R}$ and kinetic term of $\eta$ due to mixing between $S^+$ and $\eta^+$. 
Thus, we concentrate on the other phenomenology in the current paper, instead of dark matter phenomenology. 
In a different parameter set, however, it would be a promising dark matter candidate which would be discussed elsewhere.

%%%%%%%%%%%%%%%%%%%%%%%%%%%%%%%%%%%
\section*{Acknowledgments}
T.N. is supported by the Fundamental Research Funds for the Central Universities. 
H.O. is supported by Zhongyuan Talent (Talent Recruitment Series) Foreign Experts Project. 
Y.S. is supported by Natural Science Foundation of China under Grant No. W2433006. 
%%%%%%%%%%%%%%%%%%%%%%%%%%%%%%%%%%%

%\bibliographystyle{utphys}
\bibliography{ctma4.bib}
\end{document}